\documentclass[aps, pra, reprint, groupedaddress, showpacs, superscriptaddress]{revtex4-1}

\usepackage{amsmath}
\usepackage{graphicx,color}
\usepackage[colorlinks=true,linkcolor=blue, linktoc=all]{hyperref}

\def\rtHz{\sqrt{\rm Hz}}
\def\mrtHz{ \frac{{\rm m}}{\rtHz}}

\begin{document}

\newcommand{\R}[1]{\textcolor{red}{#1}}

\title{Quantum correlation measurements in interferometric gravitational wave detectors}

\author{%
\noindent
D.~V.~Martynov,$^{7}$  
V.~V.~Frolov,$^{3}$  
S.~Kandhasamy,$^{16}$ 
K.~Izumi,$^{5}$  
H.~Miao,$^{32}$	
N.~Mavalvala,$^{7}$  
E.~D.~Hall,$^{1}$  
R.~Lanza,$^{7}$	   
B.~P.~Abbott,$^{1}$  
R.~Abbott,$^{1}$
% ...
\\
\medskip
(LSC Instrument Authors)
\\
\medskip
}\noaffiliation
\affiliation {LIGO, California Institute of Technology, Pasadena, CA 91125, USA} 

\affiliation {Louisiana State University, Baton Rouge, LA 70803, USA} 

\affiliation {LIGO Livingston Observatory, Livingston, LA 70754, USA} 

\affiliation {Syracuse University, Syracuse, NY 13244, USA} 

\affiliation {LIGO Hanford Observatory, Richland, WA 99352, USA} 

\affiliation {SUPA, University of Glasgow, Glasgow G12 8QQ, United Kingdom} 

\affiliation {LIGO, Massachusetts Institute of Technology, Cambridge, MA 02139, USA} 

\affiliation {Columbia University, New York, NY 10027, USA} 

\affiliation {University of Western Australia, Crawley, Western Australia 6009, Australia} 

\affiliation {University of Florida, Gainesville, FL 32611, USA} 

\affiliation {Leibniz Universit\"at Hannover, D-30167 Hannover, Germany} 

\affiliation {Albert-Einstein-Institut, Max-Planck-Institut f\"ur Gravi\-ta\-tions\-physik, D-30167 Hannover, Germany} 

\affiliation {The University of Sheffield, Sheffield S10 2TN, United Kingdom} 

\affiliation {Stanford University, Stanford, CA 94305, USA} 

\affiliation {University of Sannio at Benevento, I-82100 Benevento, Italy and INFN, Sezione di Napoli, I-80100 Napoli, Italy} 

\affiliation {The University of Mississippi, University, MS 38677, USA} 

\affiliation {University of Michigan, Ann Arbor, MI 48109, USA} 

\affiliation {American University, Washington, D.C. 20016, USA} 

\affiliation {University of Oregon, Eugene, OR 97403, USA} 

\affiliation {University of Adelaide, Adelaide, South Australia 5005, Australia} 

\affiliation {SUPA, University of Strathclyde, Glasgow G1 1XQ, United Kingdom} 

\affiliation {Australian National University, Canberra, Australian Capital Territory 0200, Australia} 

\affiliation {University of Minnesota, Minneapolis, MN 55455, USA} 

\affiliation {Hobart and William Smith Colleges, Geneva, NY 14456, USA} 

\affiliation {Faculty of Physics, Lomonosov Moscow State University, Moscow 119991, Russia} 

\affiliation {The University of Texas Rio Grande Valley, Brownsville, TX 78520, USA} 

\affiliation {MTA E\"otv\"os University, ``Lendulet'' Astrophysics Research Group, Budapest 1117, Hungary} 

\affiliation {SUPA, University of the West of Scotland, Paisley PA1 2BE, United Kingdom} 

\affiliation {University of Maryland, College Park, MD 20742, USA} 

\affiliation {California State University Fullerton, Fullerton, CA 92831, USA} 

\affiliation {Monash University, Victoria 3800, Australia} 

\affiliation {University of Birmingham, Birmingham B15 2TT, United Kingdom} 

\affiliation {University of Washington, Seattle, WA 98195, USA} 

\date{\today}

\begin{abstract}
Quantum fluctuations in the phase and amplitude quadratures of light set limitations on the sensitivity of modern optical instruments.
The sensitivity of the interferometric gravitational wave detectors, such as the Advanced Laser Interferometer Gravitational wave Observatory (LIGO), is limited by quantum shot noise, quantum radiation pressure noise, and a set of classical noises.
We show how the quantum properties of light can be used to distinguish these noises using correlation techniques.
Particularly, in the first part of the paper we show estimations of the coating thermal noise and gas phase noise, hidden below the quantum shot noise in the Advanced LIGO sensitivity curve. 
We also make projections on the observatory sensitivity during the next science runs.
In the second part of the paper we discuss the correlation technique that reveals the quantum radiation pressure noise from the background of classical noises and shot noise.
We apply this technique to the Advanced LIGO data, collected during the first science run, and experimentally estimate the quantum correlations and quantum radiation pressure noise in the interferometer for the first time.

\end{abstract}


\maketitle

\section{Introduction}

Interferometric gravitational wave detectors have triggered extensive research in the field of quantum optics~\cite{caves_quantum_noise_1981, caves_quantum_formalism}. The standard quantum limit~\cite{Braginsky_1992, kimble_2001}, related to the Heisenberg uncertainty principle, sets limitations on the sensitivity of modern interferometric measurements.
These broadband noises, known as shot noise and quantum radiation pressure noise (QRPN), are predicted to limit the design sensitivity of Advanced LIGO in the frequency range 10\,Hz--10\,kHz~\cite{design_aligo, design_aligo_peter}.

Apart from quantum noises, the Advanced LIGO sensitivity was limited by a set of classical noises during the first observing run (O1) ~\cite{den_nb, O1_detector, den_thesis}. This run, lasting from September 2015 to January 2016, culminated in two direct observations of gravitational waves from binary black hole coalescences~\cite{first_detection, second_detection, first_detection_cbc, first_detection_cbc_results}. Further improvement of the observatory range requires more investigations into quantum and classical noises.

Since Advanced LIGO was limited by shot noise above 100\,Hz, the spectrum of classical noises is not directly observable at these frequencies. Here, we report on the use of correlation technique and reveal, for the first time, the classical-noise spectrum, hidden underneath the shot noise in Advanced LIGO. This technique explores quantum properties of light, in particular the quantum correlation among the optical power fluctuations in different readout channels. We use the obtained spectrum of classical noises to estimate the Advanced LIGO sensitivity during the next science runs, and set constraints on the coating thermal noise~\cite{Harry_2007, Evans_2008} and gas phase noise~\cite{data_gas_phase}.

In addition to estimating classical noise, we also use the correlation technique to probe QRPN in Advanced LIGO, and estimate this noise experimentally. QRPN was studied for more than thirty years ~\cite{caves_quantum_noise_1981, jaekel_1990}. It has been investigated by a number of experiments both in the gravitational wave (GW) community~\cite{corbitt_thesis, wipf_thesis, westphal_10m, khalili_12, leaver_speedmeter_2016}, and the optomechanics community~\cite{Yanbei:Review,Aspelmeyer2014, purdy_2016, clark_2016, weinstein_14, Safavi_Naeini_2013, purdy_sqz_13}. To our knowledge, its spectrum at the audio band has not yet been observed. During O1 the level of this noise is predicted to be a factor of $\simeq 8-10$ smaller compared to the current noise floor in the frequency band 30--100\,Hz; the quantum correlation, however, allows us to reveal it for the first time.

This paper is structured as follows: In Sec.~\ref{opt_conf} we discuss the configuration of the Advanced LIGO interferometers, and the propagation of the optical fields that are involved in computing the power
fluctuation of different photodiode readouts. Sec.~\ref{shot} is devoted to the investigations of the classical noise spectra in Advanced LIGO hidden below the quantum shot noise. In Sec.~\ref{qrpn} we set an experimental estimate on the level of the QRPN using the correlation technique.

\section{Optical configuration}
\label{opt_conf}

In this section, we introduce the optical configuration of the interferometer, discuss how optical fields propagate through the interferometer and beat on the photodetectors.

\subsection{The interferometer and its signal field}

\begin{figure}[ht!]
	\centering
	\includegraphics[width=0.45\textwidth]{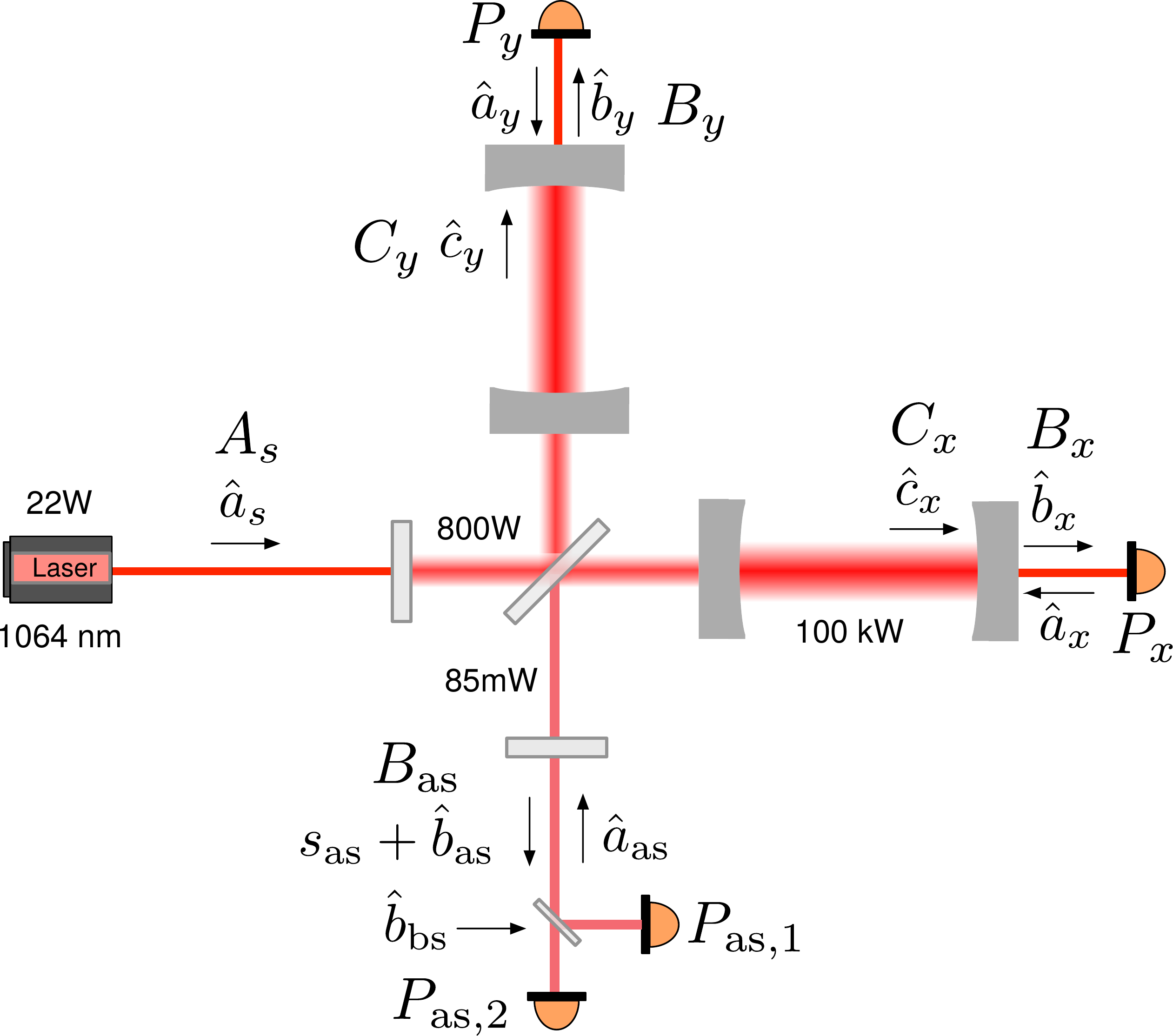}
    \caption{Layout of an Advanced LIGO detector. The annotations show the optical power in use during O1. Also shown are vacuum and laser fields entering the interferometer through the input, output, and transmission ports. GW signal $s_{\rm as}$ (also includes classical noises) leaves the interferometer through the antisymmetric port.}
	\label{fig:detector}
\end{figure}

The Advanced LIGO detectors, shown in Fig.~\ref{fig:detector}, are Michelson--type interferometers, enhanced by four optical cavities: a Fabry--P\'{e}rot cavity in each arm, one at the symmetric port and another at antisymmetric port of the interferometer~\cite{stable_cavities}. The first two arm cavities are used to optically increase the length of each arm by a factor of $G_{\rm arm}=260$. The latter two cavities are set to maximize circulating power in the interferometer by a factor of $G_{\rm prc}=38$ and optimize the frequency response to gravitational waves in the frequency range 10\,Hz -- 10\,kHz~\cite{design_aligo}, respectively. This is achieved by setting the carrier field to be anti resonant in the signal recycling cavity~\cite{Mizuno93, data_quantum_sr} and attenuating it’s power by a factor of $G_{\rm src} = 9$.

The frequency dependent GW signal is derived from the difference in the two arm lengths $L(f)$ according to the equation $L(f) / L_0$, where $L_0 = 3995$\,m is the macroscopic length of each arm and $f$ is the frequency of the GW signal. The differential arm length signal $L(f)$ is derived from the power measurement $P_{\rm as}$ at the antisymmetric port of the interferometer. The transfer function $Z = dP_{\rm as}/dL$, known as the optical response of the instrument~\cite{cal_2016, cal_s5}, can be written as
\begin{equation}\label{eq:optical_gain}
	Z(f) = \frac{4 \pi G_{\text{arm}}}{\lambda}
	\left( \frac{G_{\text{prc}} P_{\text{in}} P_{\text{as}}} {G_{\text{src}}} \right)^{1/2} K_-(f) \, \frac{\rm W}{\rm m},
\end{equation}
where $P_{\rm in}$ is the input power and $\lambda$ is the laser wavelength. The transfer function $K_- = f_- / (if + f_-)$ accounts for the diminished response of the instrument at high frequencies, where $f_-$ is known as the differential coupled cavity pole frequency~\cite{design_aligo} and is given by the equation
\begin{equation}
	f_- = \frac{T_i c}{8 \pi L_0} \approx 360 \, {\rm Hz},
\end{equation}
where $T_i \approx 0.12$ is the transmission of the signal recycling cavity and $c$ is the speed of light.

The differential arm length is sensed by using a particular type of homodyne readout technique, known as DC readout~\cite{ifo_dc_readout}. In this scheme an offset $\Delta L=10$\,pm is introduced to the differential arm length to allow a small fraction of the optical power $P_{\rm as}$ to leak to the antisymmetric port. Other longitudinal degrees of freedom are controlled using the Pound--Drever--Hall technique~\cite{design_pdh, lock_aligo_als}, with no intentional longitudinal offsets.

\begin{figure}[ht!]
	\centering
	\includegraphics[width=0.45\textwidth]{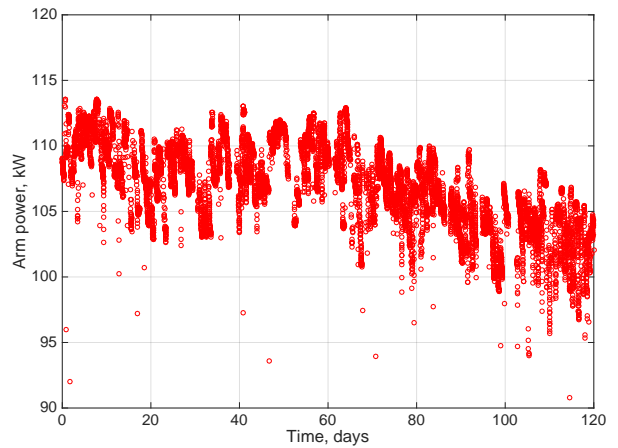}
    \caption{Optical power resonating in LIGO Livingston interferometer (L1) during the first science run. The power was fluctuating by a few percents, and slightly decreased by the end of the run due to the drift of the input power.}
	\label{fig:power}
\end{figure}

The main laser is capable of delivering 150\,W of optical power, however, only $P_{\rm in,0}=22$\,W was used during O1. This resulted in a circulating power of approximately $P_{\rm arm} = 107$\,kW in each arm. Fig.~\ref{fig:power} shows the power fluctuation in one of the arm cavities during O1. The variance was 3.2\,kW, and the precision of the power calibration was 5\%~\cite{cal_2016}. The circulating arm power has slightly decreased by the end of the run due to the drift of the input power.

\subsection{Power fluctuations as beat between DC and AC fields}

In this paper, one of the key quantities involved is the optical power fluctuation measured by different photodiodes. We treat the interferometer as a linear device in which longitudinal disturbances are linearly translated to perturbations of the optical fields. In such a linear system, the power fluctuation of a field at a frequency $f$ can be classically described as
\begin{equation}
	P(f) = A\, a^*(\nu_0-f) + A^* a(\nu_0+f),
\end{equation}
where $\nu_0=2.82 \times 10^{14}$\,Hz is the laser (carrier) frequency, $A$ is the amplitude of the carrier field at $\nu_0$ and $a(\nu_0 \pm f)$ are those of the sideband fields or perturbation fields at $\nu_0\pm f$. The superscript ``$*$'' is for complex conjugate. Similarly, when quantizing the field, the corresponding Heisenberg
operator is equal to
\begin{equation}
	\hat{P}(f) = A\, \hat{a}^\dagger(\nu_0-f) + A^* \hat{a}(\nu_0+f),
\end{equation}
where $\hat{a}$ and $\hat{a}^\dagger $ are annihilation and creation operators of the field, respectively. Throughout this paper we study physical properties of the interferometer according to the quantum formalism broadly presented in the literature~\cite{Marquardt_2007, Miao_2010, Genes_2008, lrr-2012-5}.

In Advanced LIGO, the cavity mode is excited by a laser with a large amplitude $A_s$ at an angular frequency $\omega_0=2 \pi \nu_0$. The input field is normalized according to the equation $A_s = \sqrt{P_{\rm in}/h\nu_0}$, where $h$ is the Planck constant. We study the linearized dynamics by perturbing the steady state and move into the rotating frame at $\nu_0$. Correspondingly, the carrier field is at zero frequency (DC), while the sideband fields are at frequency $\pm f$ (AC). In the following, we shall use $\hat{a}, \hat{b}, \hat{c}$, as shorthand for $\hat{a}(f), \hat{b}(f), \hat{c}(f)$ and $\hat{a}^\dagger, \hat{b}^\dagger, \hat{c}^\dagger$, as shorthand for $\hat{a}^\dagger(-f), \hat{b}^\dagger(-f), \hat{c}^\dagger(-f)$. In order to compute optical power at each particular interferometer port, one needs to calculate these two kinds of fields. The rest of this section discusses their propagation inside the interferometer.

\subsection{Propagation of the DC fields}

The static laser field $A_s$ enters the interferometer through the symmetric port and resonates in the interferometer. Optical fields in the arm cavities are denoted as $C_x$ and $C_y$ as shown in Fig.~\ref{fig:detector}. These fields then transmit to the rear side of the end mirrors (fields $B_x$ and $B_y$) and to the antisymmetric port (field $B_{\rm as}$). They are given by the equations
\begin{equation}
\begin{split}
	& C = C_x = C_y = \sqrt{\frac{1}{2} G_{\rm prc} G_{\rm arm}} A_s \\
	& B = B_x = B_y = \sqrt{T_e} C \\
	& B_{\rm as} = 2 \pi i G_{\rm arm} \frac{\Delta L}{\lambda}
	\sqrt{\frac{G_{\rm prc}}{G_{\rm src}}}  A_s,
\end{split} \label{eq:dcfields}
\end{equation}
where $T_e=3.6$\,ppm is the power transmission of the output couplers (end mirrors). The factors $G_{\rm prc}$, $G_{\rm arm}$, and 1/2 in the first equation account for the build up in the power recycling and arm cavities, and attenuation due to the 50/50 beam splitter. The factor $G_{\rm src}$ appears in the denominator of the equation for $B_{\rm as}$ since the carrier field is anti resonant in the signal recycling cavity. Note that there is a $90^\circ$ phase shift (expressed by an imaginary $i$) between the input field $A_s$ and anti symmetric port field $B_{\rm as}$ because of the transmission through the Michelson interferometer.

\subsection{Propagation of the AC fields}

Vacuum fields~\cite{caves_quantum_noise_1981, caves_quantum_formalism} enter the interferometer through the antisymmetric ($\hat{a}_{\rm as}$), symmetric ($\hat{a}_s$), and transmission ports ($\hat{a}_x$ and $\hat{a}_y$), as shown in Fig.~\ref{fig:detector}.  They propagate through the interferometer and reach the output ports according to the input--output relations~\cite{Walls_2008}. We denote output fields at the antisymmetric port as $\hat{b}_{\rm as}$, at the arm transmission ports as $\hat{b}_x$ and $\hat{b}_y$, and at the arm cavities as $\hat{c}_x$ and $\hat{c}_y$. In the case with no longitudinal offsets in the interferometric degrees of freedom ($\Delta L = 0$), and ignoring quantum radiation pressure effects, considered in Sec.~\ref{qrpn}, we can write up to the first order in $t_e=\sqrt{T_e}$ and $t_i=\sqrt{T_i}$
\begin{equation}\label{eq:vac_prop}
\left(
\begin{matrix}
	\hat{b}_{\rm as} \\
	\frac{\hat{b}_y - \hat{b}_x}{\sqrt{2}} \\
	\frac{\hat{c}_y - \hat{c}_x}{\sqrt{2}}
\end{matrix}
\right)
\simeq
 \left(
\begin{matrix}
	-K_-/K_-^*         & 0  & \sqrt{2} t_e g_- \\
	\sqrt{2} t_e g_-  & 0  & -1 \\
	\sqrt{2} g_-      & 0  & \frac{t_e}{t_i}\sqrt{2}g_- \\
\end{matrix}
\right)
\left(
\begin{matrix}
	\hat{a}_{\rm as} \\
	\hat{a}_s \\
	\frac{\hat{a}_y - \hat{a}_x}{\sqrt{2}}
\end{matrix}
\right),
\end{equation}
where $g_- = \sqrt{G_{\rm arm}/2G_{\rm src}} \times K_-(f)$. This approximation is valid for small $t_e$ and $t_i$; more precisely, energy conservation always gives $|X_{11}|^2 + |X_{12}|^2+ |X_{13}|^2 = 1$, where $X_{11}$, $X_{12}$ and $X_{13}$ are matrix elements with corresponding indices.

Eq.~\ref{eq:vac_prop} shows that the vacuum field from the laser $\hat{a}_s$ does not couple to the antisymmetric port and differential transmission signals. While an intentional offset $\Delta L$ in the differential arm length is important for accurately obtaining the DC field at the antisymmetric port~(\ref{eq:dcfields}), we find that the effect of $\Delta L$ is rather insignificant in the propagation matrix for the AC fields for Advanced LIGO. 

\section{Removing shot noise and characterizing classical noises}
\label{shot}

In this section, we describe a correlation technique for estimating the amount of classical noises buried below the shot noise.
The strength of this noise can be quantified by its spectral density $S_{\rm as}(f) = 2 P_{\rm as} h \nu$. Using Eq.~(\ref{eq:optical_gain}) we can convert this noise to the units of length. The shot noise spectrum in the GW channel $S_{\rm shot}$ limits the sensitivity of Advanced LIGO above 100\,Hz~\cite{O1_detector, den_nb} and is given by the equation
\begin{equation}\label{eq:shot_m}
	\begin{split}
	& \sqrt{S_{\rm shot}} = \sqrt{\frac{S_{\rm as}}{1-\eta}} \frac{1}{|Z(f)|} \\
	& = 2.33 \times 10^{-20} \left( \frac{\text{107\,kW}}{P_{\text{arm}}} \right)^{1/2}
	\frac{1}{|K_-(f)|} \frac{\text{m}}{\sqrt{\text{Hz}}} \\
	\end{split}
\end{equation}
where $\eta=0.28$ is the power loss from the signal recycling cavity to the photodetectors at the antisymmetric port.

The out—going field at the anti--symmetric port is split into two beams by a 50/50 beam splitter, and a homodyne detection is performed on each of the beams, as shown in Fig.~\ref{fig:detector}. In this section, we show that shot noise and photodetector dark noise can be removed, while interferometer classical and radiation pressure noises kept intact, by performing a correlation measurement between the two detectors.

\subsection{Shot and dark noise removal}

Power fluctuations at the two photodetectors at the antisymmetric port (see Fig.~\ref{fig:detector}) arise from the shot noise $P_{{\rm as},j}^{\rm shot}$, QRPN $P_{{\rm as},j}^{\rm qrpn}$, classical noises $P_{{\rm as},j}^{\rm cl}$ and photodetector dark noises $P_{{\rm as},j}^{\rm dark}$ according to the equation
\begin{equation}\label{eq:as_power_sum}
\hat{P}_{{\rm as},j} = \hat{P}_{{\rm as},j}^{\rm shot} + 
	\hat{P}_{{\rm as},j}^{\rm qrpn} + \hat{P}_{{\rm as},j}^{\rm cl} + \hat{P}_{{\rm as},j}^{\rm dark},
\end{equation}
where $j=1,2$.

Eq.~\ref{eq:as_power_sum} can be written as the beat of the static field $B_{\rm as}$ with classical field $s_{\rm as}$ and vacuum fields $\hat{b}_{\rm as}$ and $\hat{b}_{\rm bs}$. The latter field comes in through the open port of the 50/50 beam splitter in front of the photodetectors. Power fluctuations can be written as
\begin{equation}\label{eq:qu_pd_1_2}
\begin{split}
	\hat{P}_{\rm as,1} & = \frac{1}{\sqrt{2}} i B_{\rm as}^* (\hat{\nu}_{\rm as,ph} + \hat{\nu}_{\rm bs,ph} + \nu_{\rm cl,ph}) + \hat{P}_{{\rm as},1}^{\rm dark} \\
	\hat{P}_{\rm as,2} & = \frac{1}{\sqrt{2}} i B_{\rm as}^* (\hat{\nu}_{\rm as,ph} - \hat{\nu}_{\rm bs,ph} + \nu_{\rm cl,ph}) + \hat{P}_{{\rm as},2}^{\rm dark},
\end{split}
\end{equation}
where $\hat{\nu}_{\rm as,ph}$, $\hat{\nu}_{\rm bs,ph}$ and $\nu_{\rm cl,ph}$ are phase quadratures of the fields $\hat{b}_{\rm as}$, $\hat{b}_{\rm bs}$ and $s_{\rm as}$ defined as $\hat{\nu}_{\rm x, ph} = (\hat{x} - {x}^\dagger)/(\sqrt{2}i)$. Note that $\nu_{\rm cl,ph}$ includes QRPN since this noise is indistinguishable from classical noises at the antisymmetric port.

Then we compute the cross spectral density $S_{12}$ between signals $P_{\rm as,1}$ and $P_{\rm as,2}$. From Eq.~\ref{eq:qu_pd_1_2} we can write
\begin{equation}
	S_{12} = S_{\rm b,as} - S_{\rm b,bs} + \frac{1}{4}(S_{\rm cl} + S_{\rm qrpn}),
\end{equation}
where $S_{\rm b,as}$ and $S_{\rm b,bs}$ are spectra of power fluctuations due to vacuum fields $\hat{b}_{\rm as}$ and $\hat{b}_{\rm bs}$, $S_{\rm cl}$ is the spectrum of classical interferometer noises and $S_{\rm qrpn}$ is the spectrum of QRPN. Since $S_{\rm b,as} = S_{\rm b,bs}=B_{\rm as}B_{\rm as}^* h\nu/2$, the cross spectral density $S_{12}$ removes the shot noise from the GW spectrum (4 $S_{12} = S_{\rm cl} + S_{\rm qrpn}$). Note that dark noises of the photodetectors are incoherent and cancel out from $S_{12}$.

\begin{figure}[ht!]
	\centering
	\includegraphics[width=0.45\textwidth]{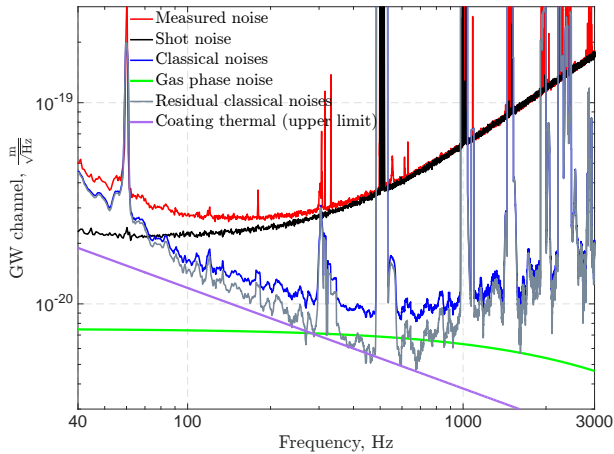}
    \caption{Noise spectra of the GW channel in the LIGO Livingston interferometer. The red and blue traces show the total noise level and the spectrum of classical noises, respectively. The black trace shows the spectrum of the shot noise. The green and gray traces show the estimated level of the gas phase noise and the sum of classical noises without this noise, respectively. The magenta trace shows the upper limit on the coating thermal noise in Advanced LIGO.}
	\label{fig:thermal}
\end{figure}

Fig.~\ref{fig:thermal} shows the calibrated cross--correlation amplitude spectrum, computed using the data from the Livingston interferometer. The spectrum of the interferometer classical noises is determined by the equation
\begin{equation}\label{eq:S_cl}
S_{\rm cl} = 4 S_{12} - S_{\rm qrpn} \approx 4 S_{12}\,,
\end{equation}
since $S_{\rm qrpn} \ll S_{\rm cl}$ in the current configuration (see Sec.~\ref{qrpn}). Above 40\,Hz this spectrum reveals the level of the classical noises in the gravitational wave channel. This result is applied to set the upper limit for the coating thermal noise~\cite{harry_coating_2012} (cf. Sec. \ref{sec:coating}) and verify the level of the gas phase noise~\cite{data_gas_phase} (cf. \ref{gas_phase}). The estimated spectrum of classical noises also provides the potential to predict the sensitivity of the Advanced LIGO detectors during future science runs (cf. Sec. \ref{sec:future}), in which shot noise will be reduced by increasing the laser power, and squeezed states of light will be introduced~\cite{caves_quantum_noise_1981, caves_quantum_formalism}.

\subsection{Coating thermal noise}\label{sec:coating}

Dielectric coatings used in the LIGO detectors consist of alternative layers of materials with low (SiO$_2$) and high (Ta$_2$O$_5$) index of refraction. Thermal noise in these coatings arises from mechanical dissipation in the coating materials, guided by the fluctuation--dissipation theorem~\cite{data_thermal_levin}. This noise is theoretically predicted to be one of the limiting noise sources for the Advanced LIGO design sensitivity in the frequency range 50\,Hz--500\,Hz~\cite{Harry_2007, Evans_2008}, as well as for the proposed next generation of the gravitational wave detectors~\cite{Evans_2016, white_paper_15}. For this reason, direct measurement of the coating thermal noise in Advanced LIGO is of significant importance.

Theoretical models depend on parameters such as the mechanical loss angles, Poisson ratio, and Young’s modulus~\cite{hong_thermal_2013}. However, due to uncertainties in the multilayer parameters, theoretical predictions have limited accuracy (up to a few tens of percent). The first table top experiment that directly measured the coating thermal noise of the Advanced LIGO coating sample predicted that the noise level is a factor of 1.22 above the theoretical prediction~\cite{Gras_2016}.

Since the coating thermal noise is coherent between the two photodetectors at the antisymmetric port, we can reveal its spectrum $S_{\rm CTN}(f)$ using the quantum correlation technique and O1 data. The estimated upper limit for this noise is
\begin{equation}
\sqrt{S_{\rm CTN}(f)} \leq 1.6 \times 10^{-19} \frac{1}{\sqrt{f}} \mrtHz.
\end{equation}
This upper limit can be improved if known classical noises are subtracted from the cross spectrum $\sqrt{S_{\rm cl}}$. Above 100\,Hz the largest contribution comes from the gas phase noise, discussed in Sec.~\ref{gas_phase}. Once this noise is incoherently subtracted, the upper limit for the coating thermal noise is
\begin{equation}
\sqrt{S_{\rm CTN}(f)} \leq 1.2 \times 10^{-19} \frac{1}{\sqrt{f}} \mrtHz.
\end{equation}
This upper limit is a factor of $\simeq 1.2$ larger than the theoretically predicted Advanced LIGO coating thermal noise~\cite{Evans_2008}.

\subsection{Gas phase noise}\label{gas_phase}

The Advanced LIGO core optics are kept under high vacuum with an average pressure of $p\simeq 1\,\mu$Pa. The presence of residual gas in the 4\,km beam tubes causes extra noise in the differential arm channel. Broadband phase noise is induced by the stochastic transit of molecules through the laser beam in the arm cavities~\cite{data_gas_phase}. This noise may limit the ultimate sensitivity that Advanced LIGO can achieve using the same vacuum infrastructure between 30\,Hz and 10\,kHz. For this reason, it is important to measure and verify the models of the gas phase noise.

The model described in~\cite{data_gas_phase} leads to the spectrum of the gas phase noise $S_{\rm gas}$ described by the equations
\begin{equation}
\begin{split}
	& \sqrt{S_{\rm gas}} = 4 \times 10^{-21}  N_{\rm gas} \mrtHz, \\
	& N_{\rm gas} = \left( \frac{\zeta_{\rm gas}}{\zeta_{H_2}} \right) \left( \frac{m_{\rm gas}}{m_{H_2}} \right) ^{1/4}
	\left( \frac{p}{10^{-6}\,{\rm Pa}} \right)^{1/2},
\end{split}
\end{equation}
where $\zeta_{\text{gas}}$ is the polarizability of the gas molecules.

\begin{figure}[ht!]
	\centering
	\includegraphics[width=0.45\textwidth]{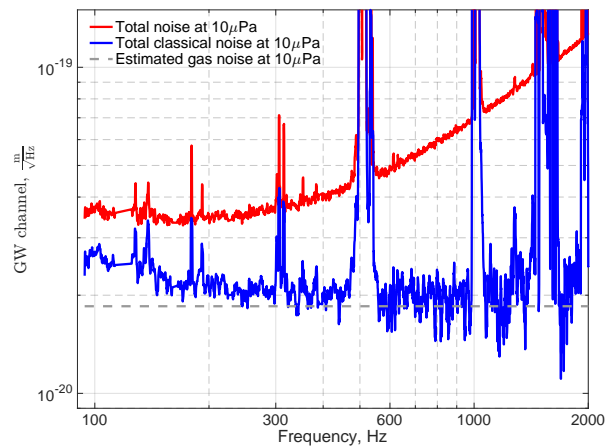}
    \caption{Measurement of the gas phase noise in the Livingston detector when the nitrogen pressure was increased up to 10$\mu$\,Pa.}
	\label{fig:gas}
\end{figure}

Gas phase noise was measured by deliberately increasing the pressure of N$_2$ in one of the arm cavities up to $10\,\mu$Pa. Under this condition, we confirmed that gas phase noise dominated over other classical noises. Fig.~\ref{fig:gas} shows the measurement of this noise under the described conditions. Even though gas phase noise is below shot noise, the quantum correlation technique has revealed its spectrum.

\subsection{Future sensitivity}\label{sec:future}

Quantum shot noise can be improved by increasing the input power as described in Eq.~(\ref{eq:shot_m}) and/or introducing squeezed states of light at the antisymmetric port~\cite{caves_quantum_noise_1981, caves_quantum_formalism}. Higher optical power is expected during the next science runs after a set of technical difficulties is solved, such as damping of the parametric instabilities~\cite{data_pi_brag, data_pi_llo}, suppressing unstable angular modes~\cite{ang_instability}, and compensation for the thermally induced wavefront distortion~\cite{aidan_tcs}. Squeezed states of light have already been demonstrated in the interferometric GW detectors~\cite{geo_squeezed, lisa_squeezed, hartmut_squeezed}. In Advanced LIGO, this technique will be enhanced by the filter cavity~\cite{Oelker_16, Isogai_2015}, alignment sensing and control~\cite{Schreiber_16} and phase control of squeezed vacuum states~\cite{Dooley_15}.

The future sensitivity of the Advanced LIGO with zero signal recycling cavity detuning~\cite{future_detune_src}, increased input power $P_{\rm in}$ and squeezed state of light can be approximated using the equation
\begin{equation}\label{eq:total_noise}
	S_{\rm gw}(P_{\rm in}) \approx S_{\rm cl} + \xi_1 S_{\rm qrpn}(P_{\rm in}) + \xi_2 S_{\rm shot}(P_{\rm in}) + S_{\rm dark},
\end{equation}
where $S_{\rm gw}$ is the power spectral density of the GW channel in units of ${\rm m}^2/\rm Hz$, $S_{\rm dark}$ is the readout electronics noise in the GW channel, and parameters $\xi_1$ and $\xi_2$ are used to define squeezing efficiency. These two parameters are enough to get an accurate estimation of the future sensitivity using the current level of classical noises. A more precise equation for the quantum noise with a filter cavity is given in~\cite{kwee_decoherence_2014}. In this section, we assume that the classical noises do not depend on the optical power resonating in the interferometer. Then Eq.~(\ref{eq:total_noise}) can be written as
\begin{equation}\label{eq:total_noise}
\begin{split}
	S_{\rm gw}(P_{\rm in}) &\approx 4 S_{12} - S_{\rm qrpn}(P_{\rm in,0})
		+ \xi_1 S_{\rm qrpn}(P_{\rm in}) + \\
		&+ \xi_2 S_{\rm shot}(P_{\rm in}) + S_{\rm dark}.
\end{split}
\end{equation}

\begin{figure}[ht!]
	\centering
	\includegraphics[width=0.45\textwidth]{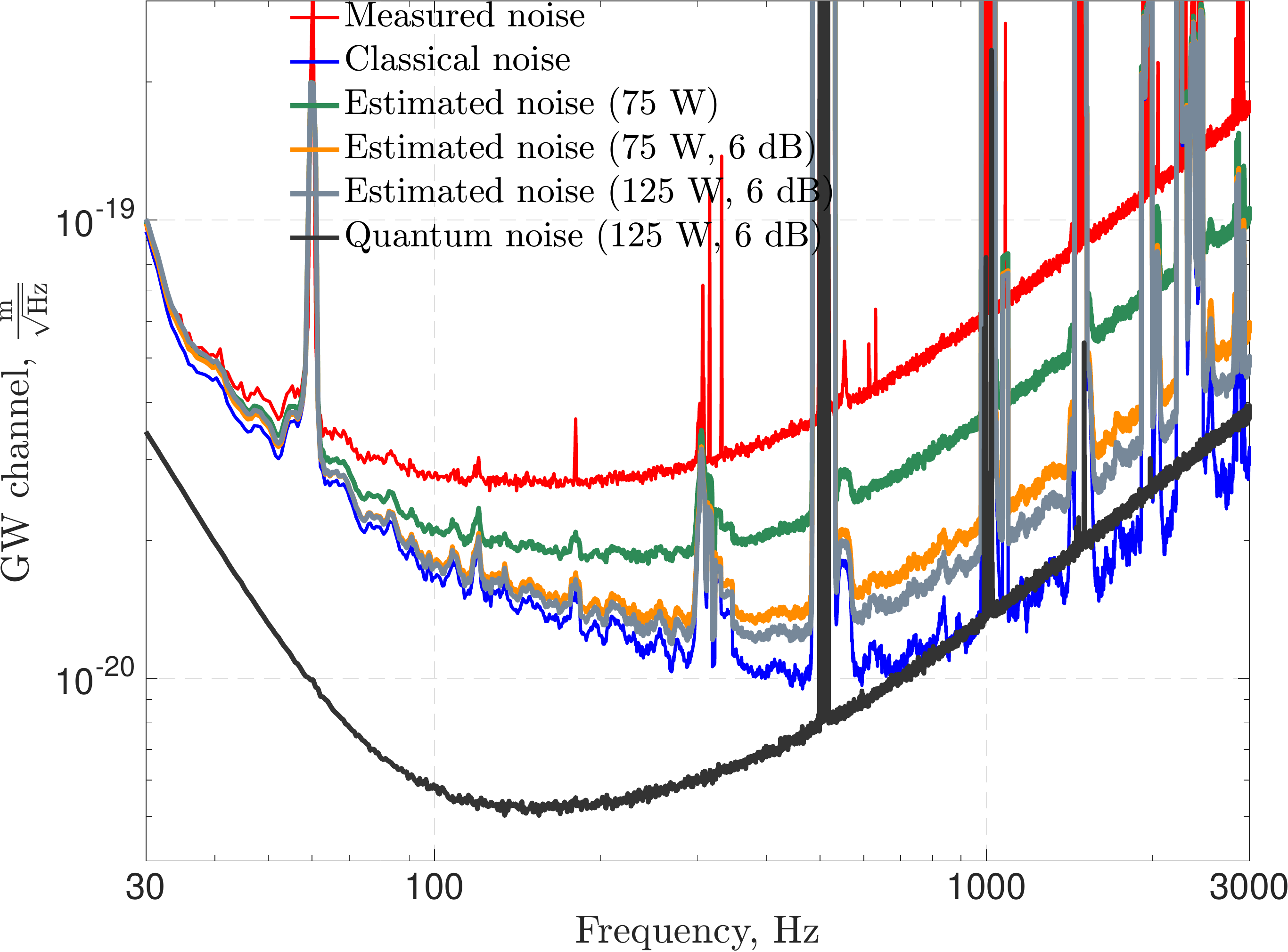}
    \caption{Estimated noise spectra of LIGO interferometers for future science runs when the input power is increased and squeezed states of light are introduced. The blue trace shows the incoherent sum of the interferometer classical noises and the readout electronics.}
	\label{fig:future_range}
\end{figure}

Fig.~\ref{fig:future_range} shows an example of the sensitivity for the input power of $P_{\rm in} = 75$\,W and $P_{\rm in} = 125$\,W. The noise spectra are computed using the current spectrum of classical noises. Since there is a significant gap between the shot noise and classical noises above 100\,Hz, Advanced LIGO sensitivity can be significantly improved using squeezed states of light. We assume that the shot noise is reduced by 6\,dB using the squeezing technique ($\xi_2 = 0.25$). Recent study on the realistic filter cavities with optical losses~\cite{matt_realistic_13} shows that QRPN does not significantly improve or degrade when squeezed states of light are introduced ($\xi_1 \approx 1$). In this section we assume that QRPN is not effected by squeezing.

\begin{table}[h!]
    \caption{Estimated observatory range with the current classical noise.}\label{table:range}
    \begin{ruledtabular}
    \begin{tabular}{c c c c}
    Input power,\,W & Squeezing,\,dB & NS--NS,\,Mpc & BH--BH,\,Mpc \\ \hline
	22 & 0 & 95 & 841 \\
	75 & 0 & 116 & 894 \\
	75 & 6 & 136 & 932 \\
	125 & 6 & 148 & 916
    \end{tabular}
    \end{ruledtabular}
\end{table}

Table~\ref{table:range} summarizes the projected Advanced LIGO range for merging binary neutron stars (NS--NS) with 1.4 solar masses each and binary black holes (BH--BH) with 30 solar masses each. This range is defined by the distance, averaged over the sky location and source orientation, at which the binary coalescence can be detected with SNR of 8. Corresponding noise curves are shown in Fig.~\ref{fig:future_range}.

Table~\ref{table:range} shows that once the input power $P_{\rm in}$ is increased from 75\,W up to 125\,W with 6\,dB of squeezing, the observatory sensitivity to $(1.4+1.4)M_\odot$ neutron stars is improved due to shot noise reduction. However, the sensitivity to $(30+30)M_\odot$ black holes is reduced due to the increase of QRPN. This noise is the main object of study in the next section.

\section{Quantum radiation pressure}
\label{qrpn}

In this section, we discuss correlation technique that reveals quantum correlation in the Michelson--type interferometers due to QPRN. The theoretically calculated strength of this noise during O1 in the GW channel is given by the equation~\cite{den_nb}
\begin{equation}\label{eq:qrp_darm}
\begin{split}
	\sqrt{S_{\rm qrpn}} &= \beta \frac{1}{f^2} |K_-(f)| \frac{\text{m}}{\sqrt{\text{Hz}}} \\
	\beta &\simeq \frac{2}{\pi^2 cM} \sqrt{\frac{P_{\rm arm}G_{\rm arm}h \nu_0}{G_{\rm src}}} \\
	&= (1.32 \pm 0.06) \times 10^{-17} \frac{\text{m Hz}^2}{\sqrt{\text{Hz}}}.
\end{split}
\end{equation}

During O1 the spectrum of QRPN was a factor of $\sqrt{S_{\rm gw}/S_{\rm qrpn}}\simeq 8-10$
smaller than the total sum of noises, given by Eq.(\ref{eq:total_noise}), in the frequency range 30--100\,Hz. However, QRPN is predicted to be a limiting noise source below 50\,Hz once the design sensitivity is achieved~\cite{design_aligo}. For this reason, it is important to study QRPN experimentally.

\subsection{Quantum correlations}


In Michelson--type interferometers it is possible to reveal QRPN by making a cross--correlation measurement of the differential transmission signal $\Delta P_{\rm tr} = P_y - P_x$ with the GW channel. The main idea is that both channels are correlated due to the amplitude quadrature of the vacuum fields $\hat{a}_{\rm as}$, $\hat{a}_x$, and $\hat{a}_y$, defined as $\hat{\nu}_{\rm x, a} = (\hat{x} + \hat{x}^\dagger)/2$. A similar approach was already considered for single Fabry--P\'{e}rot cavities~\cite{Perdu_2013, Verlot_2009}.

\begin{figure*}[t!]
	\centering
	\begin{minipage}{0.495\textwidth}
	\includegraphics[width=0.99\textwidth]{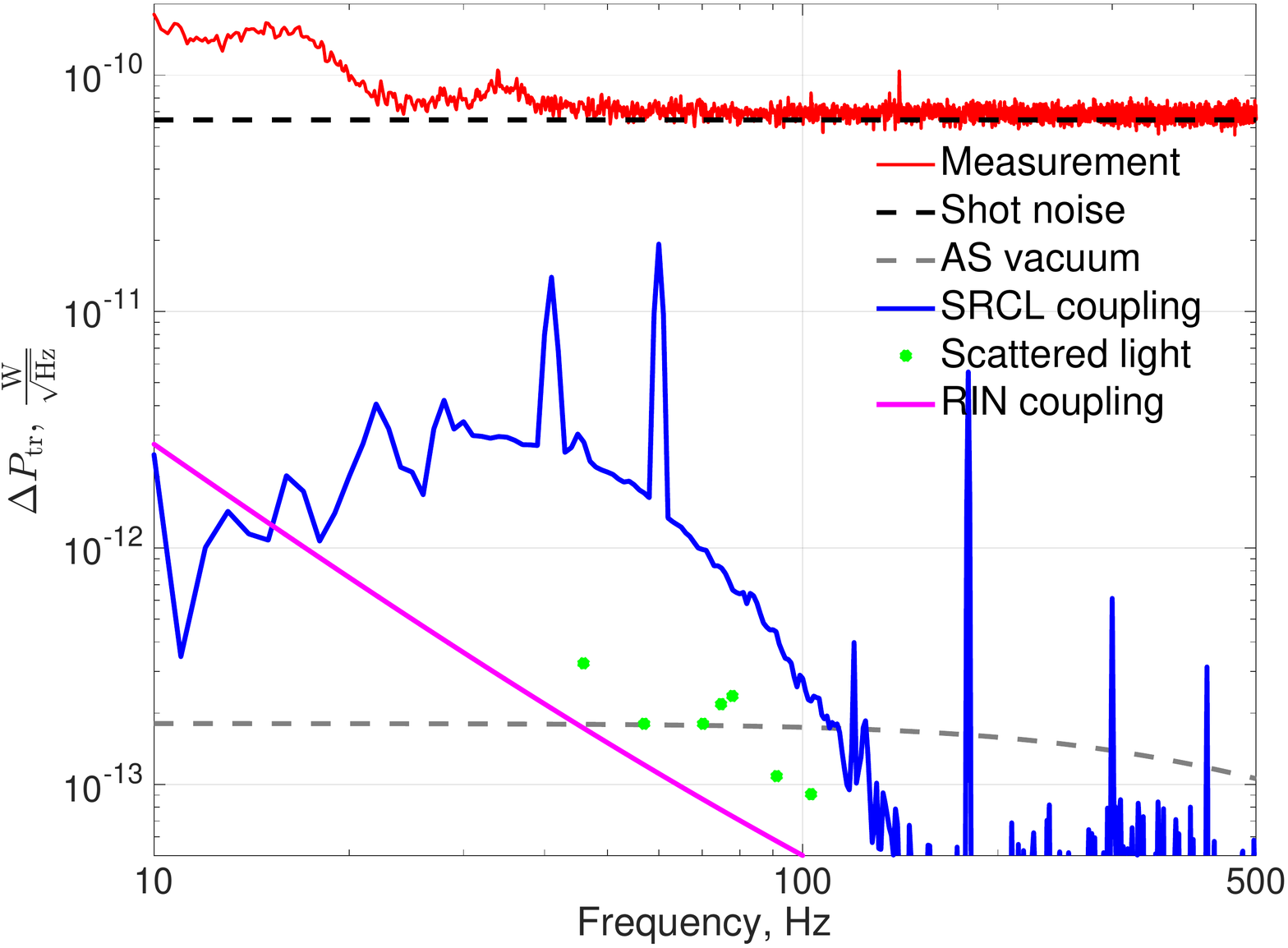}
    	\end{minipage}
	\begin{minipage}{0.495\textwidth}
        \includegraphics[width = 0.99\textwidth]{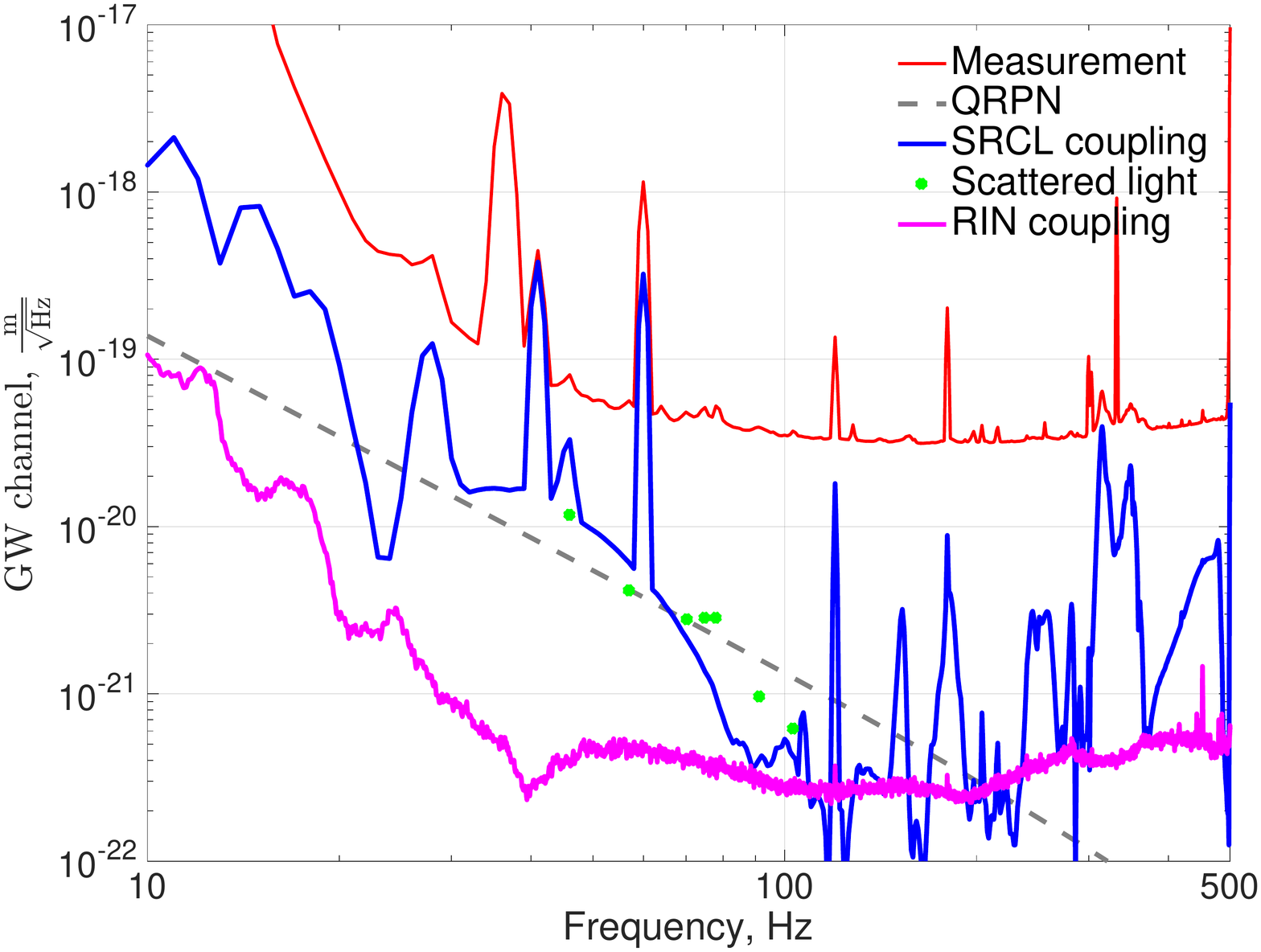}
    	\end{minipage}
	\\ \vspace{2mm}
    \begin{minipage}{0.495\textwidth}
        \mbox{(a) Differential arm transmission signal.}
    \end{minipage}
    \begin{minipage}{0.495\textwidth}
        \mbox{(b)  GW channel.}
    \end{minipage}
	\caption{Coupling of classical and quantum noises to the differential transmission channel (shown on the left)and GW channel (shown on the right). Blue curves show the coupling of the signal recycling cavity length (SRCL) fluctuations. Magenta traces show the coupling of the relative intensity fluctuations (RIN).}
	\label{fig:tr_gw_nb}
\end{figure*}

In order to see quantum correlations between the GW channel and $\Delta P_{\rm tr}$, we derive QRPN and $\Delta P_{\rm tr}$ as a function of vacuum fields $\hat{a}_{\rm as}$ and $\hat{a}_x - \hat{a}_y$.

Quantum power fluctuations in the inline (perpendicular) arm cavity are caused by the beating of the static field $C_x$ ($C_y$) against the vacuum field $\hat{c}_x$ ($\hat{c}_y$). Fluctuations in the circulating cavity power result in a fluctuating force on the mirrors. The QRPN is given in terms of a change in the differential arm length by
\begin{equation}\label{eq:qrp_fields}
	\hat{L}_{\rm qrpn}  = \frac{4 C (\hat{c}_y + \hat{c}^\dagger_y - \hat{c}_x - \hat{c}^\dagger_x)}{c M(2 \pi f)^2}
\end{equation}
where $M=40$\,kg is the mass of the arm cavity mirrors, considered to be a free mass above $f \geq 10$~Hz.
We can rewrite Eq.~(\ref{eq:qrp_fields}) at frequencies below the differential coupled cavity pole ($f \leq f_-$) in the form
\begin{equation}\label{eq:qrp_as_tr}
	\hat{L}_{\rm qrpn} = \frac{\beta}{h \nu_0 f^2}
	 \left( \hat{\nu}_{\rm as,a} + \frac{t_e g_-}{\sqrt{2}} \hat{\nu}_{\rm tr,a} \right),
\end{equation}
where $\hat{\nu}_{\rm as,a}$ and $\hat{\nu}_{\rm tr,a}$ are the amplitude quadratures of the fields $\hat{a}_{\rm as}$ and $\hat{a}_{\rm tr} = (\hat{a}_y - \hat{a}_x)/\sqrt{2}$, respectively.

Quantum power fluctuations measured by the differential transmission signal $\Delta P_{\rm tr}$ are caused by the vacuum fields $\hat{a}_{\rm as}$ and $\hat{a}_{\rm tr}$. Below the coupled cavity pole frequency ($f \leq f_-$) we can write
\begin{equation}\label{eq:as_tr}
\begin{split}
	\Delta \hat{P}_{\rm tr} & = B (\hat{b}_y + \hat{b}_y^\dagger - \hat{b}_x - \hat{b}_x^\dagger) \\
	& \simeq 2 \sqrt{\frac{P_{\rm arm} T_e}{h\nu_0}} \left( \sqrt{2}t_e g_-\hat{\nu}_{\rm as,a} - \hat{\nu}_{\rm tr,a} \right).
\end{split}
\end{equation}

Eqs.~(\ref{eq:qrp_as_tr}) and~(\ref{eq:as_tr}) show that the GW channel and the differential arm transmission channel are correlated due to vacuum fields. The cross--spectral density between these two signals $S_{\rm gw,tr}$ is given by the equation
\begin{equation}\label{eq:cross_qrp_tr}
	S_{\rm gw,tr}(f) = \alpha \frac{1}{f^2} \frac{\rm m\cdot W}{\rm Hz},
\end{equation}
where the coefficient $\alpha$ is determined by the optical configuration of the instrument. Using Eqs.~(\ref{eq:qrp_as_tr}) and~(\ref{eq:as_tr}) we can write
\begin{equation}\label{eq:alpha_eq}
	\alpha = \frac{2}{\pi^2} \frac{G_{\rm arm}}{G_{\rm src}}
	\frac{h \nu_0 P_{\rm arm}}{c M} T_e T_{\rm tr},
\end{equation}
where $T_{\rm tr} \simeq 0.02$ is the power transmission from the end mirror to the photodetectors in the transmission ports. The largest uncertainties come from the absolute power stored in the arm cavities (see Fig.~\ref{fig:power}) and the power on the transmission photodetectors. The theoretical value of $\alpha$ for O1 configuration using Eq.~(\ref{eq:alpha_eq}) is
\begin{equation}\label{eq:alpha_theory}
\alpha = (8.3 \pm 0.8) \times 10^{-31} \; \rm m \cdot W \cdot Hz.
\end{equation}
Note that since $\alpha$ is a real number, the quantum correlation signal should be in the real part of the cross power spectrum ${\rm Re}[S_{\rm gw,tr}]$. 


\subsection{Classical noises}

Eq.~(\ref{eq:cross_qrp_tr}) shows that the GW channel and differential transmission signal are correlated through the quantum noises. However, any classical noise, which couples to both channels, also adds correlations. In the case of no longitudinal offsets and no imbalances in the arm cavities, classical displacement noises and laser noises do not couple to $\Delta P_{\rm tr}$. For this reason, ideally, classical noises should not add correlations between $\Delta P_{\rm tr}$ and $L$.

However, due to wanted and unwanted imbalances in the interferometer the coupling coefficient of a particular set of classical noises is non-zero. Fig.~\ref{fig:tr_gw_nb} shows the coupling of quantum and classical noises to the differential transmission and GW channels. The dominant classical couplings are discussed below in this section.

\subsubsection{Displacement noises}

First, we consider displacement noise in the arm cavities $L_{\rm disp}$ measured by the GW channel $L$. For the differential arm offset of $\Delta L = 10$\,pm, the coupling coefficient of differential arm signal to transmission signals is simulated as
\begin{equation}\label{eq:disp}
\Delta P_{\rm tr} \simeq \frac{10^4}{f} \left[ \frac{\rm W}{\rm m} \right] \times L_{\rm disp}
\end{equation}
in the frequency range 10\,Hz to 1\,kHz. According to Eq.~(\ref{eq:disp}), power fluctuations measured at the transmission ports due to the residual differential arm fluctuations are $\Delta P \sim 10^{-15} \text{\,W}/\sqrt{\text{Hz}}$, in the frequency range from 30\,Hz to 100\,Hz. Since this number is 5 orders of magnitude below the shot noise level, correlations between $\Delta P_{\rm tr}$ and $P_{\rm as}$ due to displacement noises in the arm cavities are insignificant.

\subsubsection{Laser noises}

Relative laser intensity noise (RIN) couples to $\Delta P_{\rm tr}$ due to unwanted imbalances $Q$ between the two arms according to the equation
\begin{equation}\label{eq:rin}
	\frac{\Delta P_{\rm tr}(f)}{\Delta P_{\rm tr}(0)} = Q \frac{1}{K_+(f)} \text{RIN}(f),
\end{equation}
where the transfer function $K_+ = f_+ / (if + f_+)$ accounts for the diminished response of the interferometer to the common motion of the two arms, and $f_+ = 0.6$\,Hz.
The imbalances ($Q \sim 1\%$) are caused by different optical losses and photodetector responses. During O1 the input intensity noise was suppressed to RIN$\sim 10^{-8}-10^{-7}/\sqrt{\text{Hz}}$. This noise also directly couples to the GW channel due to the DC readout technique, as shown in Fig.~\ref{fig:tr_gw_nb}. Eq.~(\ref{eq:rin}) shows that RIN is passively filtered at 0.6\,Hz by the common cavity pole. These reasons make classical intensity noise insignificant in the correlation measurement of $\Delta P_{\rm tr}$ and $P_{\rm as}$.

\subsubsection{Scattered light}

A small fraction of light is scattered out from the main beam due to coating roughness~\cite{loss_arm_scatter, scatter_peter_hiro, ripple_peter, design_aligo_peter}. This light hits the walls of the vacuum chambers, becomes modulated in phase, and is backscattered to the main beam. The phase quadrature of the scattered light is measured as sensing noise at the GW channel, but is not detected in the arm transmission channels. The amplitude quadrature is mixed with the main beam and actuates on the interferometer mirrors similar to the QRPN. Since the amplitude quadrature of the scattered light noise is also detected in the arm transmission channels, this noise introduces classical correlations between $\Delta P_{\rm tr}$ and $L$.

\begin{table}[h!]
    \caption{Scattered light peaks.}\label{table:peaks}
    \begin{ruledtabular}
    \begin{tabular}{c c c}
    Frequency\,[Hz] & $L_{\rm ph}$\,[m] & $U$\,[$\sqrt{\rm Hz}$] \\ \hline
	46 & $5.5 \times 10^{-20}$ & 2.2  \\
	57 & $2.5 \times 10^{-20}$ & 1  \\
	70 & $2.5 \times 10^{-20}$ & 1  \\
	75 & $3.1 \times 10^{-20}$ & 1.2  \\
	78 & $3.3 \times 10^{-20}$ & 1.3  \\
	91 & $1.5 \times 10^{-20}$ & 0.6  \\
	103 & $1.3 \times 10^{-20}$ & 0.5
    \end{tabular}
    \end{ruledtabular}
\end{table}

Phase modulation of the scattered field is determined by the motion of the scattering object. Since the vacuum chambers are not seismically isolated, the RMS of their motion is $\sim 1$\,$\mu$m, with a spectral density of $\sim 1$\,nm$/\sqrt{\rm Hz}$ around 30--300\,Hz. Scattering locations have not been identified up to this moment, and we can estimate classical correlations in $S_{\rm gw,tr}$ only based on the measurement of the phase quadrature of the scattered light in the GW channel.

Table~\ref{table:peaks} summarizes the acoustic peaks present in the GW channel (see Fig.~\ref{fig:thermal}). $L_{\rm ph}$ is the RMS of the peak, where the subscript ``ph'' emphasizes that this noise was produced by phase modulation of the scattered light. $U$ determines the ratio between the RMS of the scattering peak and the shot noise level. Since we assume that the power of scattered light in the phase and amplitude quadratures is the same, then $U$ also determines the ratio of the radiation pressure force on the mirrors due to scattered light to QRPN.

Using the ratio $U$ we can estimate the contribution of the scattered light amplitude quadrature to the GW channel and differential transmission signal. These couplings are shown in Fig.~\ref{fig:tr_gw_nb}.

\subsubsection{Signal recycling cavity length}

Next we consider fluctuations in the signal recycling cavity length (SRCL), $L_{\text{src}}$. Due to the differential arm offset $\Delta L$, these fluctuations couple to the GW channel through the classical radiation pressure noise according to the equation
\begin{equation}
	L(f) = \frac{0.16}{f^2} \left[ \frac{\rm m}{\rm m} \right] \frac{\Delta L}{10\text{\,pm}} L_{\rm src}(f),
\end{equation}
and to the differential arm transmission signal according to the equation
\begin{equation}
	\Delta P_{\rm tr} = 10^5 \left[ \frac{\rm W}{\rm m} \right] \frac{\Delta L}{10\text{\,pm}} L_{\rm src}(f).
\end{equation}
Online feedforward cancellation is used to subtract SRCL from the GW channel. This system reduces the coupling by a factor of 5 to 10. Residual correlations are subtracted during the post processing scheme described below.

\subsubsection{Feedforward cancellation}

Once measured by the separate sensors, classical noises can be canceled out from the analysis. We computed cross power spectra of SRCL with the GW channel $S_{\rm gw,src}$ and with the differential transmission signal $S_{\rm src,tr}$. We then computed the cross--spectrum between the GW and transmission channels $S_{\rm gw,tr}$ using the equation
\begin{equation}\label{eq:qrp_ff}
	S_{\rm gw,tr} = S_{\rm gw,tr}^{(0)} -
		\frac{S_{\rm gw,src}S_{\rm src,tr}}{S_{\rm src}},
\end{equation}
where $S_{\rm src}$ is the power spectral density of the signal recycling cavity length fluctuations and $S_{\rm gw,tr}^{(0)}$ is the raw computed cross--spectrum between the GW and transmission channels before we apply the feedforward cancellation scheme.


\subsection{Analysis of O1 data}

\begin{figure}[ht!]
        \centering
        \includegraphics[width=6cm, angle=270]{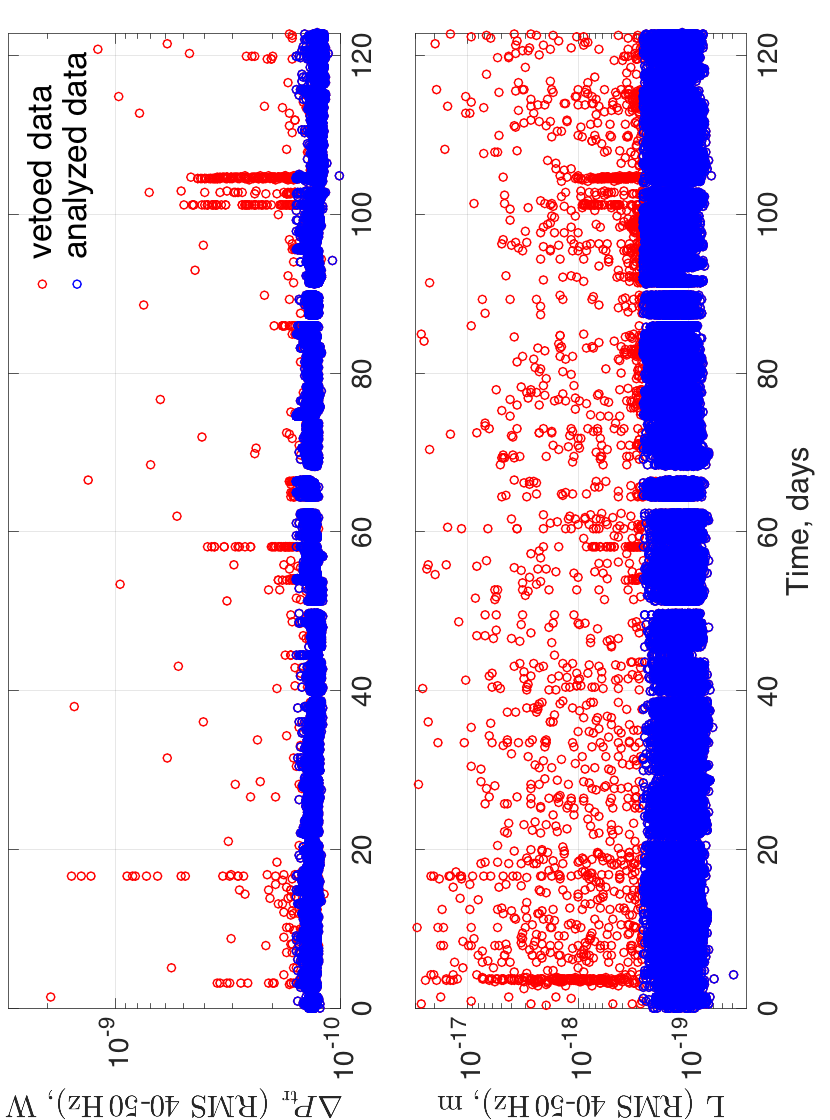}
        \caption{Data quality cut. The quality of the data is estimated based on the RMS of the signal computed between 40 and 50\,Hz. Red points are subtracted from the analysis.}
        \label{fig:DQcut}
\end{figure}

\begin{figure}[ht!]
	\centering
	\includegraphics[width=0.45\textwidth]{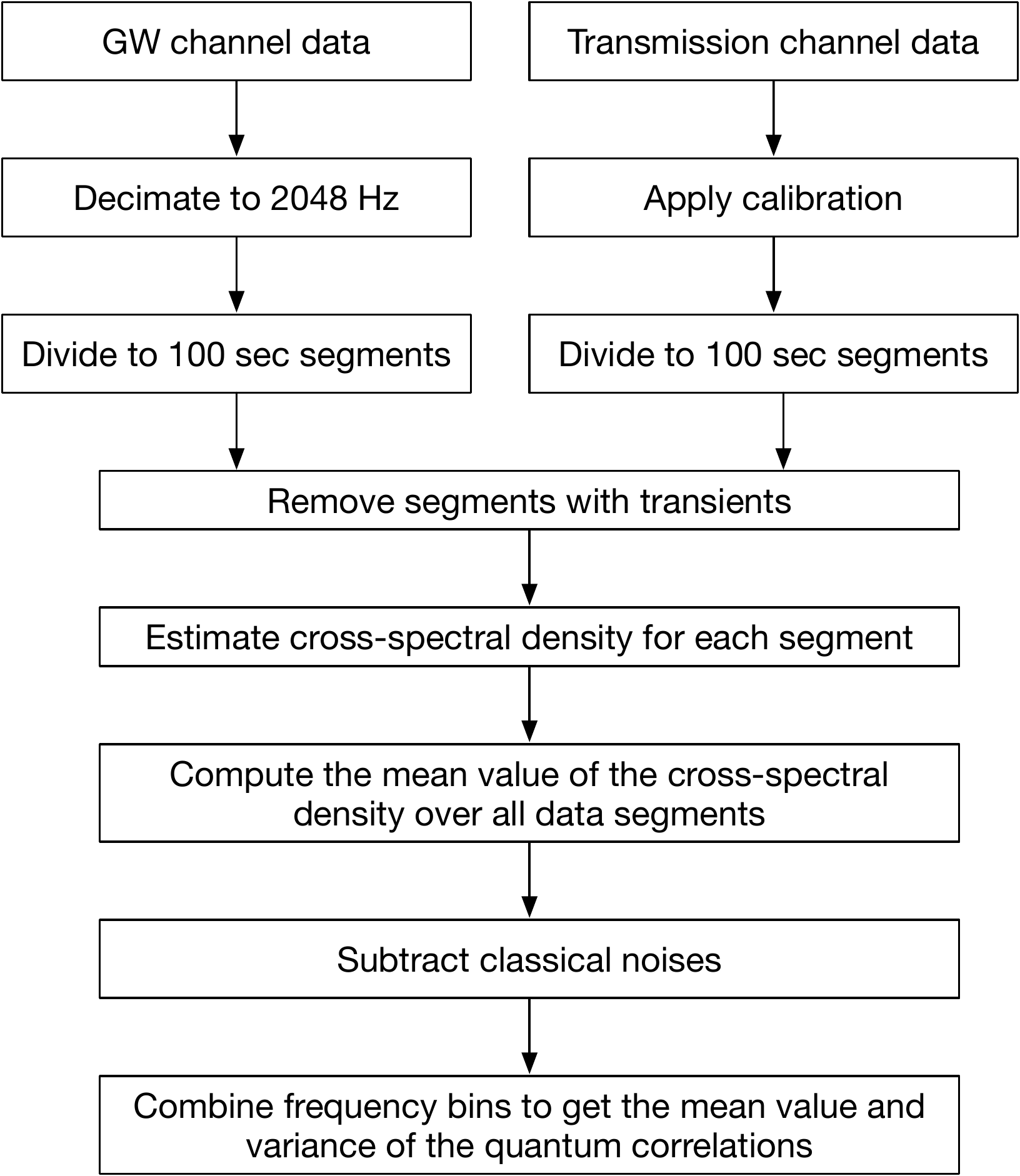}
    	\caption{Data analysis flow diagram for the estimate of the quantum correlations.}
	\label{fig:pipeline}
\end{figure}

We started with O1 data collected from the LIGO Hanford interferometer (H1) when the detector was in the linear regime, in an undisturbed state. A flow diagram of the data analysis pipeline is shown
in Fig.~\ref{fig:pipeline}. The GW channel was decimated from 16384\,Hz down to 2048\,Hz to match the sampling rate of the transmission channel. We then divided those data into 100\,sec segments and calculated power spectral densities $S_{\rm gw}$, $S_{\rm tr}$ and the cross-spectral density $S_{\rm gw,tr}$. We used 1-sec, 50 \% overlapping, and Hann windowed FFTs for these estimations. We also high-passed the individual signals at 10\,Hz to reduce spectral leakage due to the windowing.

Even though we used data segments from the interferometer's undisturbed state, changes in the environment such as high seismic motion, dust particles, and electronics failures could produce short transients in the data. In order to remove such transients, we applied a cut based on deviations of the RMS of the power spectral densities in the 41-50 Hz band. Fig.~\ref{fig:DQcut} shows an example of applying such a cut. The final spectra were produced by combining power spectrum and cross -- spectral densities from individual 100\,sec intervals that passed the above cut. After the cut, we were left with $N \simeq 4.7 \times 10^6$ 1-sec data segments ($\simeq$1300 hours of data).

Once the data was cleaned of glitches, we applied the feedforward cancellation scheme [Eq.~(\ref{eq:qrp_ff})] to subtract the signal recycling cavity noise from $S_{\rm gw,tr}$. The total amount of remaining data sets a statistical limit on the estimate of the cross power spectrum $S_{\rm gw,tr}(f)$. At each frequency bin the variance of the noise is given by the equation
\begin{equation}\label{eq:stat_lit}
	\sigma^2(f) = \frac{S_{\rm gw}(f) S_{\rm tr}(f)}{2N},
\end{equation}
where factor of 2 accounts for the fact that the quantum signal is in the real quadrature of $S_{\rm gw,tr}$ according to Eq.~(\ref{eq:cross_qrp_tr}).

\begin{figure}[ht!]
	\centering
	\includegraphics[width=0.45\textwidth]{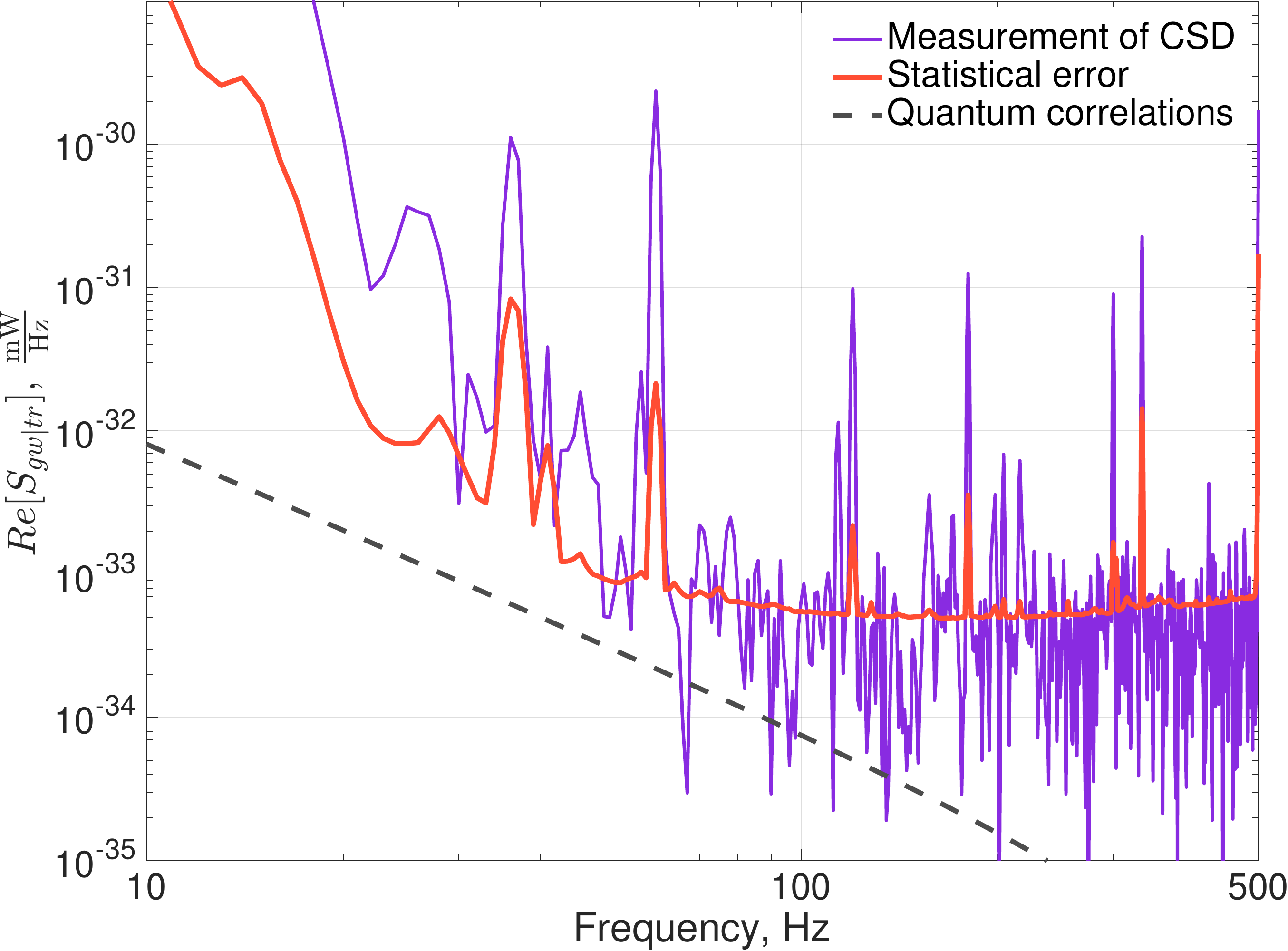}
    	\caption{Cross--spectral density between the GW channel and the differential transmission channel computed using O1 data. The red trace shows the square--root of the variance at each frequency bin. The dashed line shows the mean value of the theoretically predicted quantum correlations.}
	\label{fig:csd}
\end{figure}

The statistical limit $\sigma(f)$ from Eq.~(\ref{eq:stat_lit}) and the real part of $S_{\rm gw,tr}(f)$ are shown in Fig.~\ref{fig:csd}, where the bandwidth of the cross--spectral density is $\Delta f = 1$\,Hz. Above 30\,Hz the cross--spectral density is limited by the statistical limit and classical lines. These lines are subtracted from the analysis using the following condition: if the height of the line is larger than $5\sigma$, then this frequency bin is not used in the computation of the signal--to--noise ratio.

We combined measurements at different frequencies from 30\,Hz to 300\,Hz in order to get the final estimation of the QRPN~\cite{allen_1999, ligo_stochastic_2004}. The experimental value of $\alpha$ from Eq.~(\ref{eq:cross_qrp_tr}) is given by the equation
\begin{equation}\label{eq:alpha_est}
	\alpha_{\rm est} = \frac{\sum {\rm Re}[S_{\rm gw,tr}] f^2 \zeta}{\sum \zeta}
		= 10.8 \times 10^{-31} \; \rm m \cdot W \cdot Hz,
\end{equation}
where $\zeta = 1/(\sigma^2 f^4)$ and we sum over all frequency bins included in the analysis. Using this result, we can also set an experimental estimation on QRPN by using the equation
\begin{equation}\label{eq:qrpn_exp}
	\beta_{\rm est} = \sqrt{\frac{2 \alpha_{\rm est}}{\pi^2 M c T_e T_{\rm tr}}} 
	= 1.57 \times 10^{-17} \frac{\text{m Hz}^2}{\sqrt{\text{Hz}}}.
\end{equation}

The statistical variance of $\alpha_{\rm est}$ is given by the equation
\begin{equation}\label{eq:sigma}
	\sigma_{\alpha, {\rm stat}} =   \frac{1}{\sqrt{\sum \zeta}}
	 = 4.5 \times 10^{-31} \; \rm m \cdot W \cdot Hz.
\end{equation}

Apart from the statistical error, we also calculate uncertainties due to calibration precision~\cite{cal_2016} and imbalance of the impedances in the arm transmission photodetectors
\begin{equation}\label{eq:sigma_alpha_cal}
\begin{split}
	\sigma_{\alpha, {\rm cal}} &= 1.1 \times 10^{-31} \; \rm m \cdot W \cdot Hz \\
	\sigma_{\alpha, {\rm imb}} &= 3.9 \times 10^{-31} \; \rm m \cdot W \cdot Hz.
\end{split}
\end{equation}

The uncertainty on $\beta_{\rm est}$ comes from the uncertainties on $\alpha_{\rm est}$, given in Eqs.(\ref{eq:sigma},\ref{eq:sigma_alpha_cal}) and $T_{\rm tr}$. They sum incoherently and the total variance of $\beta_{\rm est}$ is given by the equation
\begin{equation}
	\sigma_{\beta} = 0.45 \times 10^{-17} \frac{\text{m Hz}^2}{\sqrt{\text{Hz}}}.
\end{equation}
Eqs.(\ref{eq:alpha_est},~\ref{eq:qrpn_exp}) and (\ref{eq:alpha_theory},~\ref{eq:qrp_darm}) show that the measured quantum correlation and QRPN coefficients $\alpha_{\rm est}$ and $\beta_{\rm est}$ are consistent to the theoretically predicted values $\alpha$ and $\beta$ within the error bars. The SNR can be improved by collecting more data, increasing the input power, or reducing classical noises in the GW channel as discussed in the next section.

\subsection{Prospects for the future runs}

The estimate of the quantum correlation coefficient, $\alpha_{\rm est}$, can be improved by reducing the statistical error, $\sigma_{\alpha}$, and the classically induced correlations between the GW and transmission channels, $\alpha_{\rm cl}$. In this section we describe how the estimate of quantum correlations can be further improved.

In order to improve the SNR of the estimation (reduce $\sigma_{\alpha} $), we need to collect more data. Other parameters might also change during future runs. From Eqs.~(\ref{eq:alpha_est}) and~(\ref{eq:sigma}) we can write
\begin{equation}\label{eq:snr}
	{\rm SNR} \propto \sqrt{\frac{t P_{\rm arm} T_{\rm tr}}{S_{\rm gw}}},
\end{equation}
where $t$ is the integration time. Eq.~(\ref{eq:snr}) shows that in order to improve the SNR by a factor of 2, we need to increase either the integration time or $P_{\rm arm}$ or $T_{\rm tr}$ by a factor of 4. Alternatively, we need to reduce classical noises in the GW channel by a factor of 2.

At the same time, we need to reduce $\alpha_{\rm cl}$ for better estimation of the quantum correlations. The signal recycling cavity length noise was already subtracted during the analyses presented in this paper, and classical correlations between $L$ and $\Delta P_{\rm tr}$ should be much less compared to the quantum correlations. Cancellation of the other noises, such as scattered light noise and possibly unknown noises, can reduce $\alpha_{\rm cl}$ even further. 

\section{Conclusions}
\label{conclusions}

We have applied the correlation technique, which explores quantum properties of light, to reveal both classical and quantum noise spectra underlying the observed sensitivity curve of Advanced LIGO. Particularly, in the first part, we estimated the spectrum of the classical noises during O1, taking into account that shot noise is not correlated between the two photodetectors at the antisymmetric port. Using this spectrum we set an upper limit on the coating thermal noise~\cite{Harry_2007, Evans_2008}. We estimated future detector sensitivity when the input power is increased and quantum shot noise is reduced. We also verified the model of the gas phase noise~\cite{data_gas_phase} by modulating the pressure in one of the Advanced LIGO beam tubes.

In the second part of the paper we estimated the QRPN using O1 data. Theoretical calculations show that the GW channel and the arm transmission channels should be coherent through the quantum back action noise. We experimentally estimated quantum correlation and QRPN during O1. Our results are consistent with the theoretically predicted values within the error bars. The SNR can be improved during the subsequent observing runs. This approach is also helpful for other experiments limited by quantum noises, such as the MIT PDE experiment~\cite{wipf_thesis}, the AEI 10\,m prototype~\cite{westphal_10m}, Virgo~\cite{intro_virgo}, and KAGRA~\cite{intro_kagra}.

\section*{Acknowledgments}

The authors gratefully acknowledge the support of the United States National Science Foundation (NSF) and the Kavli Foundation. LIGO was constructed by the California Institute of Technology and Massachusetts Institute of Technology with funding from the NSF, and operates under cooperative agreement PHY-0757058. Advanced LIGO was built under award PHY-0823459. This paper carries LIGO Document Number LIGO-P1600280.

\bibliography{paper}

\end{document}